\documentstyle[12pt]{article}
\begin{document}
\title{Self-Similar Collapse in Brans-Dicke Theory and Critical Behavior}
\author{H. P. de Oliveira}
\maketitle
\label{address}
Universidade do Estado do Rio de Janeiro, Instituto de F\'{i}sica, R.
  S\~{a}o Francisco Xavier, 524, Maracan\~{a}, CEP 20550-013, Rio de Janeiro.
Brazil. E-mail: henrique@vmesa.uerj.br

\begin{abstract}
We use the technique of conformal transformations to generate
self-similar collapse in Brans-Dicke theory. We analyze the
solutions concerning the critical behavior found recently by Choptuik.
The critical exponent associated to the formation of black hole for near
critical evolution is obtained. The role of the coupling parameter is
discussed.
\end{abstract}
PACS number: xxx
\vfill\eject

\hspace\parindent

Recently, Choptuik\cite{chop} has showed numerically the occurrence of
critical phenomena in spherically symmetric gravitational collapse of massless
scalar field. These critical phenomena are characterized by power-law
behavior, scaling relations and a form of universality. In his study, Choptuik
analyzed the evolution one-parameter families of initial conditions (parameter
$p$, say), and classified the solutions in three classes: (i)subcritical
($p<<p_{cr}$), where the final state is a complete dispersion of scalar field
resulting in flat spacetime; (ii) supercritical ($p>>p_{cr}$) in which the
scalar field is intense enough to trap itself in a black hole; (iii)critical
solutions ($p=p_{cr}$), that represents the transition between complete
dispersal and black hole formation. The most interesting results merge for
near critical ($p \cong p_{cr}$) evolutions due to two important features.
First, the mass of formed black hole in this regime is given by $M_{BH}
\propto |p-p_{cr}|^\gamma$, where $\gamma \cong 0.37$ is conjectured to be
universal since it does not depend on the particular initial condition.
Second, for near critical evolution a strong field region exists where the
scalar field oscillates in a very particular way. Such oscillations are echoes
and the scalar field satisfies $\phi(e^{-n\,\Delta}\,t,e^{-n\,\Delta}\,r)
\cong \phi(t,r)$, where is an integer number and $\Delta \cong 3.4$. It is
important to mention that the exact critical solution has an infinite train of
echoes in the strong field region.

Quite surprisingly the same results were obtained for axisymmetric collapse of
gravitational waves\cite{ae} and for the spherically symmetric collapse of
radiative fluids\cite{evans} exhibiting local self-similarity. Therefore, it
is strongly suggested that these critical phenomena are independent of the
symmetries as well as the collapsing matter. Analytical solutions describing
such features could be extremely useful for their understanding.
Unfortunately, this task has proved to be very difficult, and an alternative
approach was to suppose continuous self-similarity. In several self-similar
exact solutions with ordinary scalar field the critical exponent was found to
be 0.5\cite{brady}, whereas for the conformally coupled case,
0.21\cite{hened}. Maison\cite{maison}, analyzing the collapse of perfect fluid,
showed the non-universality of the critical exponent in the sense that it
depends strongly on the state equation relating the energy density and
pressure.

In this paper, our objective is to study analytically the self-similar
collapse in Brans-Dicke theory of gravitation ($BDT$), and its relation with
the critical behavior as described above. We are motivated by the renewed
interest in the called scalar-tensor\cite{varios} theories of gravitation in
which the $BDT$ is the most simple. We mention, for instance, the numerical
work performed by Scheel et al\cite{schell} concerning the collapse of
matter in $BDT$. We will consider here a simplified model where the matter
field is absent. Then, we expect to find, if it is present, the role of the
coupling parameter $\omega$ on the value of the critical exponent.

$BDT$ is the most simple of the scalar-tensor theories. In such theories the
spacetime is characterized by the metric tensor $g_{\mu\nu}$ and by a scalar
field coupled to the metric and matter. The action for the $BDT$ is:

\begin{equation}
S=\int\,d^{4}\,x\,\sqrt{-g}\,(\phi\,R-\frac{\omega}{\phi}\,g^{\mu\nu}\,
\partial_{\mu}\,\phi\,\partial_{\nu}\,\phi+L_m)
\end{equation}

\noindent where $\phi$ is the scalar field, $\omega$ the coupling constant and
$L_m$ is the matter Lagrangian. For a more general scalar-tensor theory
$\omega=\omega(\phi)$. The basic characteristic of this theory is the variable
gravitational term measured locally as
$G=\frac{2\,\omega+4}{\phi\,(2\,\omega+3)}$. According with the most recent
experimental tests, $\omega$ has minimum value about 500. As showed by
Dicke\cite{dicke}, it is possible to express the theory in the Einstein frame.
In this frame the action (1) is formally given as General Relativity with
ordinary scalar field, which is a suitable redefinition of the scalar field
$\phi$. Then, we have:

\begin{equation}
S=\int\,d^{4}\,x\,\sqrt{-\tilde{g}}\,(\tilde{R}-\tilde{g}^{\mu\nu}\,
\partial_{\mu}\,\Phi\,\partial_{\nu}\,\Phi)
\end{equation}

\noindent where $\tilde{R}$ is the scalar of curvature obtained from the
metric $\tilde{g}_{\mu\nu}$, $\Phi$ is the ordinary scalar field and we have
considered $L_m=0$. Both actions integrals are equivalent if the following
conformal transformation holds:

\begin{equation}
g_{\mu\nu}=\frac{1}{\phi}\,\tilde{g}_{\mu\nu},
\end{equation}

\noindent and the scalar field $\Phi$ is expressed in terms of $\phi$ by:

\begin{equation}
\Phi=\sqrt{\frac{2\,\omega+3}{2}}\,ln\,\phi
\end{equation}

\noindent Therefore, from the above equations, it is possible to generate
solutions in BDT starting from those solutions for ordinary scalar fields in
GR.

To generate solutions in $BDT$ describing self-similar spacetimes we
consider the Roberts'\cite{roberts} solution. Such a solution has been studied
as an analytical model that exhibits critical behavior in scalar field
collapse\cite{brady}. The Roberts' solution is given by:

\begin{equation}
d\,s^2=-d\,u\,d\,v+r^2(u,v)\,(d\,\theta^2+sin^2\,\theta\,d\,\varphi^2)
\end{equation}

\noindent where $u$ and $v$ are null coordinates, and

\begin{equation}
r^2(u,v)=-\frac{u\,v}{2}+a_2\,u^2+b_2\,v^2=\frac{1}{16\,a_2}\,f_+\,
(u,v)f_-\,(u,v)
\end{equation}

\noindent with:

\begin{equation}
f_{\pm}\,(u,v)=4\,a_2\,u-(1 \pm \sqrt{1-16\,a_2\,b_2})\,v.
\end{equation}

\noindent The scalar field $\Phi$ is:

\begin{equation}
\Phi=\pm\frac{1}{\sqrt{2}}\,ln\left|\frac{f_+(u,v)}{f_-(u,v)}\right|
\end{equation}

\noindent The collapse situation is properly described if $a_2$ is chosen
positive, more specifically, $a_2=1/4$, without loss of generality. The
solutions for which $a_2<0$ corresponds to expanding self-similar
cosmological models. The collapse solutions are classified according to
three distinct values of the parameter $b_2$: $0<b_2<1/4$, the subcritical
regime, $b_2=0$, the critical regime and the supercritical $b_2<0$. In this
last case, there is the formation of a "black hole" type configuration (Fig.
1(a)), meaning that the formed singularity is enclosed by an apparent
horizon, but the spacetime is not asymptotically flat. The critical exponent
related to near critical evolution was found to be equal to 0.5.

To obtain the spherically symmetric self-similar collapse in $BDT$, we begin
with Eqs. (4) and (8), which yields:

\begin{equation}
\phi(u,v)=\left(\pm\,\frac{f_+(u,v)}{f_-(u,v)} \right)^{\beta(\omega)}
\end{equation}

\noindent where $\beta(\omega)=\frac{\pm\,1}{\sqrt{2\,\omega+3}}$.
According with Eqs. (3) and (9), we have:

\begin{equation}
d\,s^2=-\phi^{-1}\,d\,u\,d\,v+\Sigma^2(u,v)\,(d\,\theta^2+sin^2\,
\theta\,d\,\varphi^2)
\end{equation}

\noindent where the proper area, $\Sigma^2(u,v)$, is given by:
\begin{equation}
\Sigma^2(u,v)=\frac{1}{16\,a_2}\,\phi(u,v)^{-1}\,f_-(u,v)\,f_+(u,v).
\end{equation}

\noindent The conformal factor, $\phi$, is regular everywhere with exception
of $f_-(u,v)=0$ or $f_+(u,v)=0$, depending of the sign of $\beta(\omega)$.
Indeed these regions represent, in general, singularities\footnote{For
instance, the scalar of curvature is given by
$R=16\,\omega\,\beta^2\,(1-4\,b_2)\,u\,v\,\left(\frac{\pm\,f_-^{1-2/\beta}}
{f_+^{1+2/\beta}}\right)$. Since $\omega>0$, $|\beta|<1$ meaning that at
$f_-(u,v)=0$ or $f_+(u,v)=0$, the scalar of curvature diverges. For sake of
completeness, only if $|\beta|>2$ (or $-1.5<\omega<-1.375$) $R$ is finite at
$f_-(u,v)=0$ ($\beta<0$) and $f_+(u,v)=0$ ($\beta>0$).}. As an important
property of conformal transformations, the causal structure of the spacetime
is not changed if the transformation is well behaved, which is the present
case. We assume $\omega>0$, implying $|\beta(\omega)|<1$ and, as a
consequence, the singular regions have zero proper area. Otherwise, for
$\omega<0$ ($|\beta(\omega)|>1)$, the singular regions could be characterized
by an infinity proper area. Two types of solution emerge from Eq. (9): the one
with positive sign inside the parenthesis, and another with negative sign. For
the first type, we must have $\frac{f_+(u,v)}{f_-(u,v)}>0$. Hence, we chose
$a_2=1/4$ so that the proper area is given by
$\Sigma(u,v)^2=\frac{1}{4}\,f_-(u,v)^{1+\beta(\omega)}\,
f_+(u,v)^{1-\beta(\omega)}$.
For the second type, it is necessary that $\frac{f_+(u,v)}{f_-(u,v)}<0$.
Consequently, $a_2$ must be negative in order the proper area to be positive
as required. However, due to the fact the collapse is not described for such a
choice of $a_2$, we are not going to study this situation. Nonetheless, $a_2$
can be positive only if $|\beta(\omega)|$ is an even number that is realized
only by $\omega<0$.

The apparent horizon is determined locally by:

\begin{equation}
g^{\mu\nu}\,\Sigma_{,\mu}\,\Sigma_{,\nu}=0.
\end{equation}

\noindent After substituting Eq. (11) into (12), two relations are obtained:
\begin{equation}
u_{AH}=(1+\beta(\omega)\,\sqrt{1-4\,b_2})\,v
\end{equation}

\begin{equation}
u_{AH}=\frac{4\,b_2\,v}{1-\beta(\omega)\,\sqrt{1-4\,b_2}}
\end{equation}

\noindent where the first equation corresponds to $\Sigma_{,u}=0$ whereas the
second $\Sigma_{,v}=0$. Note that in the limit $\omega \rightarrow \infty$, we
recover the results provided by Roberts' solution as expected. The next step
is to verify for which cases the apparent horizon is present.

As mentioned, the Penrose diagrams are not altered by the conformal
transformation. Therefore, following ref. \cite{brady}, we classify the
solutions as subcritical ($0<b_2<1/4$), critical ($b_2=0$) and supercritical
($b_2<0$). Studying the presence of the apparent horizon in the collapse in
$BDT$ the following results are obtained. For $b_2=0$, it is not difficult to
show that there is no apparent horizon described either by Eq. (13) or Eq.
(14), unless $\omega<-1$ that has no interest. The subcritical case
($0<b_2<1/4$) displays some novelty. Taking
$\beta(\omega)=\frac{-1}{\sqrt{2\,\omega+3}}<0$, and after a straight
analysis, we conclude that there is no apparent horizon. However, for
$\beta(\omega)=\frac{1}{\sqrt{2\,\omega+3}}>0$, the apparent horizon given by
Eq. (14) exists for any value of $\omega>0$ and encloses the timelike
singularity. This situation is also found for the collapse of conformally
coupled scalar field\cite{hened}, but has no relevance regarding the critical
behavior.

The most interesting case is the supercritical ($b_2<0$). Let us consider
first $\beta(\omega)>0$, where, after easy verification, the apparent
horizon indicated by the line $OH$ (Fig. 1(b)) is spacelike and described if
by Eq. (14) with the condition $\omega > \omega_*=-2\,b_2-1$ holds. For
$\omega \rightarrow \infty$ we recover the apparent horizon given by
Roberts' solution (line $OR$, Fig. 1(b)). On the other hand if
$\omega=\omega_*$, the horizon becomes the null surface $v=0$. In Fig. 2, we
depict the effect of varying $\omega$ on the apparent horizon. Finally,
considering $\beta(\omega)<0$ the apparent horizon is again described by Eq.
(14), but with no restriction on possible values of $\omega$. The effect of
varying $\omega$ from $-3/2$ to $\infty$ is to rotate the line $OH$ from the
$v$-axis to the line $OR$ (the apparent horizon of Roberts' solution) as
indicated in Fig. 2.

\vspace{1.0 cm}
{\bf Black Hole mass and the Power Law for Near Critical Behavior}
\vspace{0.5 cm}

We are going to see the influence of the coupling parameter in the power
law associated to the mass of the formed black hole. This last quantity is
the effective gravitational mass inside the apparent horizon:

\begin{equation}
m_{AH}=\frac{1}{2}\,\sqrt{f_-(u,v)^{1+\beta(\omega)}\,
f_+(u,v)^{1-\beta(\omega)}}\,|_{AH}
\end{equation}

\noindent Substituting Eq. (14) with $b_2<0$ into the above equation and
considering near critical evolution, $b_2 \cong 0$, we have:

\begin{equation}
m_{AH} \cong
\frac{(1-\beta(\omega))^{\frac{1+\beta(\omega)}{2}}}{1-\beta(\omega)}\,
[-(3-\beta(\omega))\,b_2+1-\beta(\omega)]^{\frac{1-\beta(\omega)}{2}}
\,(-b_2)^{\frac{1+\beta(\omega)}{2}}\,v.
\end{equation}

\noindent As mentioned previously, self-similar spacetimes fails to be
asymptotically flat and black hole configurations have infinite mass when $v
\rightarrow \infty$. However, an important point to be stressed is the
obtained power-law $m_{AH} \cong (-b_2)^\gamma$ for near critical evolution.
Notice that the exponent $\gamma$ depends strongly on $\omega$, or, in another
word, of the intensity of which the scalar field couples to the geometry.
Then, as showed in Fig. 3, if $\beta(\omega)>0$, we have $0.79<\gamma<0.5$ for
$\omega$ varying from 0 to $\infty$. In the case of $\beta(\omega)<0$,
$0.21<\gamma<0.5$ for $\omega$ varying from 0 to $\infty$. Incidentally, the
exact value 0.37 is achieved in this case when $\omega \cong 5.8$. A large
value of the coupling constant compatible with earlier experimental results,
$\omega \cong 500$, produces $\gamma=0.52$ and $0.48$ for $\beta(\omega)$
positive and negative, respectively.

We can argue that the above result as well as the one obtained for conformally
coupled scalar field rise doubts regarding the universality of the value 0.37.
In these models, however, there a serious weakness which is the assumption of
continuous self-similarity, and contrary to the problem treated by Choptuik,
the spacetime is not asymptotically flat. Koike and Mishima\cite{km} studying
the collapse of a thin massive shell coupled with an outgoing null fluid
resulting in an asymptotically flat model showed that the critical exponent is
not universal. On the other hand, from the large amount of works dealing with
self-similar collapse in connection to cosmic censorship hypothesis, a cutoff
is introduced to add an asymptotically flat region. This procedure is
equivalent to truncate the spacetime at some, say, $v=v_0$. Hence, we can
match the self-similar spacetime with Schwarzschild or outgoing Vaidya
solutions, where in the later case it models the radiation that escapes during
the collapse. In general, a null shell characterized by mass and surface
pressure\cite{Israel} would be present for such a junction . In this way, the
local trapped mass, represented by the mass function at the apparent horizon,
is given by the mass of the shell and the Brans-Dicke scalar field mass. A
reasonable condition to be imposed is that the shell has negligible mass and
pressure, which could be seem as natural from the physical point of view. If
this hypothesis is satisfied, Eq. (16) is rewritten as:

\begin{equation} m_{AH} \cong
\frac{(1-\beta(\omega))^{\frac{1+\beta(\omega)}{2}}}{1-\beta(\omega)}\,
[-(3-\beta(\omega))\,b_2+1-\beta(\omega)]^{\frac{1-\beta(\omega)}{2}}
\,(-b_2)^{\frac{1+\beta(\omega)}{2}}\,v_0. \end{equation}

\noindent We expect that, depending on the asymptotically flat spacetime
to be joined, the conditions $m_{shell} \cong 0$ and $P_{shell} \cong 0$ at
the apparent horizon, render some relation between $v_0$ and $b_2$. Then, it
will be possible to estimate, in this simplified model, the value of the
critical exponent. We are investigating these issues and the results will be
given elsewhere\cite{henwang}.

The author acknowledges the financial support of the
Brazilian agency CNPq (Conselho Nacional de Desenvolvimento Cient\'{i}fico e
Tecnol\'{o}gico).

\begin{itemize}

\item Fig. 1(a)
"Black hole" type configuration. The spacetime is not asymptotically flat and
the mass of the black hole grows without bound as $v \rightarrow \infty$. At
$v=0$ we match smoothly the scalar field solution with the Minkowski
 spacetime. The lines indicated by $R$ and $H$ represent the apparent
horizon of the Roberts and the Brans-Dicke solutions, repectively.

\item Fig. 1(b)
The lines $OR$ and $OH$ represent the apparent horizon for the Brans-Dicke
(case $\beta(\omega)>0$) and Roberts solutions, respectively. The effect of
varying $\omega$ from $\omega_*$ to $\infty$ is to rotate $OH$ in the
counterclockwise sense. For $\omega=\infty$ both lines coincide.

\item Fig. 2
For the case $\beta(\omega)<0$, the apparent horizon behaves in a different
manner as compared to the previous case. As far as $\omega$ grows, the line
$OH$ rotates in the clockwise sense and tends to $OR$ for $\omega \rightarrow
\infty$.

\item Fig. 3
Behavior of the critical exponent $\gamma(\omega)=\frac{1+\beta(\omega)}{2}$.

\end{itemize}

\end{document}